\documentclass{article}
\usepackage{spconf}

\usepackage{cite}
\usepackage{amsmath,amssymb,amsfonts}
\usepackage{algorithmic}
\usepackage[ruled,vlined, linesnumbered]{algorithm2e}
\usepackage{graphicx}
\usepackage{textcomp}
\usepackage{bm}
\usepackage{xcolor}

\def\BibTeX{{\rm B\kern-.05em{\sc i\kern-.025em b}\kern-.08em
    T\kern-.1667em\lower.7ex\hbox{E}\kern-.125emX}}
    
\newtheorem{theorem}{Theorem}

\title{Measure Concentration on the OFDM-based Random Access Channel}

\begin{document}

\name{Gerhard Wunder$^{1}$, Axel Flinth$^{2}$, Benedikt Gro\ss$^{1}$ \thanks{This work was partially funded by DFG SPP 1798 - Compressed Sensing in Information Processing. F. acknowledges support from the Chalmers AI Research center (CHAIR).}}
\address{\small 1. Department of Computer Science, 
Freie Universit\"at Berlin,
Berlin, Germany \\ \small
2. Institution for Electrical Engineering, Chalmers University of Technology, 
Gothenburg, Sweden. \\}

\maketitle

\begin{abstract}
It is well known that CS can boost massive random access protocols. Usually,
the protocols operate in some overloaded regime where the sparsity can be
exploited. In this paper, we consider a different approach by taking an
orthogonal FFT base, subdivide its image into appropriate sub-channels and let
each subchannel take only a fraction of the load. To show that this approach
can actually achieve the full capacity we provide i) new concentration
inequalities, and ii) devise a sparsity capture effect, i.e where the
sub-division can be driven such that the activity in each each sub-channel is
sparse by design. We show by simulations that the system is scalable resulting in a
coarsely 30-fold capacity increase.

\end{abstract}


\section{Introduction}

There is meanwhile an unmanageable body of literature on CS for (massive)
random access in 5G-6G networks, often termed as compressive random access
\cite{Wunder2015_ACCESS}. A comprehensive overview of competitive approaches
for 5G (Rel. 16 upwards) can be found in \cite{Bockelm2018_ACCESS}. Often
unnoticed outside the mathematical literature is that the main driving
principle of CS is \emph{measure concentration}: Let $A\in\mathbb{C}^{m\times
n}$ be a randomly subsampled FFT matrix such that $\mathbb{E}\left\Vert
Ax\right\Vert ^{2}=\left\Vert x\right\Vert ^{2}$ for any vector non-random
$x\in\mathbb{C}^{n}$, with $|\operatorname{supp}\left(  x\right)  |\leq k$,
then \cite[Lemma 12.25]{Rahut2013}:%
\begin{align}
&  \mathbb{P}\left(  \left\vert \left\Vert Ax\right\Vert ^{2}-\left\Vert
x\right\Vert ^{2}\right\vert >\left(  \sqrt{\frac{k}{m}}+\epsilon\right)
\left\Vert x\right\Vert ^{2}\right)  \nonumber\\
&  \leq2e^{-\frac{\epsilon^{2}m}{2k}\frac{1}{1+2\sqrt{k/m}+\epsilon/3}%
}\label{eq:Mconc4}%
\end{align}

This has been widely exploited for massive random access. Taking this
principle forward, in this paper we prove an extended version and show that
it can be used to design powerful massive random access systems.

The very recent papers by Choi \cite{Choi2018_TVC}\cite{Choi2020_IoT} have
been brought to our attention and have revived our interest in the problem. In
\cite{Choi2018_TVC} a two-stage, grant-free approach has been presented where
in stage one a classical CS-based detector detects the active $n$-dimensional
pilots from a large set of size $l$ which is followed in stage two by data
transmission using the pilots as spreading sequences. \cite{Choi2020_IoT} has
presented an improved version where the data slots are granted through prior
feedback. The throughput is analyzed and simulation show signficant
improvement over multi-channel ALOHA. However, missed detection analysis is
carried out under overly optimistic assumptions such as ML detection making
the results fragile (e.g. the missed detection cannot be independent of $l$ as
the results in \cite{Choi2020_IoT} suggest). Moreover no concrete pilot design
(just random)\ and no frequency diversity is considered which is crucial for
the applicability of the design.

We take a different approach here:\ Instead of overloading $n$ subcarriers
with $l>n$ resources we use $n$-point FFT (orthogonal basis) and subdivide the
available bandwidth into sub-channelss, and apply hierarchical algorithms for
the detection in each of the sub-channels and slots. Then, we show that in a
scalable system, i.e. with growing $n$, this approach yields provably the full
capacity. The proof is based on two main ingredients. First, a new measure
concentration result for a certain family of vectors with common sparsity
pattern is proved. Second, we devise a \emph{sparsity capture effect}, by coarsely
bundling (at most) $\log(n)$ frequency resources in an $n$ resource system
which are then collaboratively (CS-based) detected. With this scheme, the collision probability in each
sub-channel will diminish over $n$. In the simulation section, this
is validated for several setting system yielding an 30-fold increase in user capacity.


{\it{Notations}}.
The elements of a vector/sequence $x$ are
referred to as $(x)_{i}$. The vector $x_{\mathcal{A}}\in\mathbb{C}%
^{|\mathcal{A}|}$ (matrix $X_{\mathcal{A}}$) is the projection of elements
(rows) of the vector $x\in\mathbb{C}^{n}$ (matrix $X$) onto the subspace
indexed by $\mathcal{A}\subset\lbrack n]$.
Depending on the context we also denote by $x_{\mathcal{A}}$ the vector that coincides with $x$ for the elements indexed by $\mathcal{A}$ and is zero otherwise.

\section{System Model}

\label{sec:system}

Joint detection problems in massive random access, say in 5G uplink, can typically
be cast as follows: We allow for a fixed maximum set of $u$ users in a system
with a signal space of total dimension $n$, which can possibly be very large,
e.g. $2^{14}$ \cite{Wunder2015_GC}. The $l$-th (time domain) signature
$p_{l}\in\mathbb{C}^{n}$ of the $k$-th user is randomly taken from a possibly
large pool $\mathcal{P}\subset\mathbb{C}^{n}$ with $r$ ressources. The random
mapping $k\hookrightarrow l$ is not one-to-one obviously, so there are
collisions, and the access point "sees" a superposed channel called effective
channel of multiple users. While there is no clear way to detect such
collisions right away, there is typically some mechanism to detect such
collisons later in the collision resolution phase \cite{dera}.
After the
pilot phase the users send data over $t$ slots using the same signature

Let $h_{k}^{\prime}\in\mathbb{C}^{s}$ denote the sampled channel impulse
response (CIR) of the $k$-th user, where $s\ll n$ is the length of the cyclic
prefix. The $l$-th effective channel is defined through $h_{l}=\sum
_{k:k\hookrightarrow l}h_{k},l=1,...,r$. Furthermore, we define the matrix
$\text{circ}^{(s)}(p_{l})\in\mathbb{C}^{n\times s}$ to be the circulant matrix
with $p_{l}$ in its first column and $s-1$ shifted versions in the remaining columns.

The $k_{s}>0$ \emph{non-zero} complex-valued channel coefficients are
independent {normallly distributed} with power $\mathbb{E}|(h_{k})_{l}%
|^{2}=\sigma_{h,l}^{2}$. While $k_{u}>0$ active users have a non-vanishing
CIR, inactive users are modeled by $h_{k}^{\prime}=0$.\ Note that we assume
from now on, if not otherwise explicitly stated, that the channel energy is
equally distributed within the coefficients, which however does not affect the
generality of the results.
Hence, we
shall set without loss of generality $\sigma_{h}^{2}=1$ so that the
Signal-to-Noise Ratio ($\operatorname{SNR}$) becomes%
\[
\operatorname{SNR}:=\mathbb{E}|h_{0}|^{2}/\sigma^{2}=1/\sigma^{2}.
\]
Note, that $\operatorname{SNR}$ does not reflect the true receive
$\operatorname{SNR}$ in the system, which is $k_{s}/\sigma^{2}$.


Stacking the CIRs into a single column vector $h=[h_{1}^{T}\ h_{2}^{T}\ldots
h_{r}^{T}]^{T}$, the signal received by the base station in Stage-1 is given
by:%
\[
\text{\textbf{Stage-1: \ \ }}y_{0}=C(p)h+e_0,
\]
where $C(p)=[\operatorname{circ}^{(s)}(p_{1}),\dots,\operatorname{circ}%
^{(s)}(p_{r})]\in\mathbb{C}^{n\times rs}$ depends on the stacked signatures
$p=[p_{1}^{T}\ p_{2}^{T}\ldots p_{r}^{T}]^{T}$ and $e_0\in\mathbb{C}^{n}$ is 
AWGN with $e_0\sim\mathcal{CN}(0,\sigma^{2}I_{n})$.

In Stage-2 each active user takes the same signature and applies data in the
remaining $t-1$ slots. Let the data be collected in the vectors $d_{i}%
\in\mathbb{C}^{u},i=1,...,t-1$. For ease of exposition we assume binary or
QPSK data such that $|(d_{i})_{l}|=1$. Define the matrix $D_{i}=\mathrm{diag}%
(d_{i})\otimes I_{s}$ so that the transmission is given by:
\[
\text{\textbf{Stage-2: \ \ }}y_{i}=C(p)D_{i}h+e_{i}%
\]
Notably $h$ and $D_{i}h$ have the same non-zero locations, i.e. same support.
To have a reference point for the load of the system we will set $r=u$ without
loss of generality (except for the overloading case where $r<u$ in the
numerical section). We will refer to the percentage $k_{u}/n$ as the \emph{load
of the system.}.

A key idea in compressive random access is that the user identification and
channel estimation task needs to be accomplished within a much smaller
subspace, compared to the signal space, so that the remaining dimensions can
be exploited. The measurements in this subspace are of the form:
\begin{equation}
\left(  y_{i}\right)  _{\mathcal{B}}=\Phi_{\mathcal{B}}y_{i}=\Phi
_{\mathcal{B}}C(p)D_{i}h+\Phi_{\mathcal{B}}e_{i},\label{eq:SysMod}%
\end{equation}
where we denote the restriction of some measurement matrix to a set of rows
with indices in $\mathcal{B}\subset\lbrack n]$ by $\Phi_{\mathcal{B}}$. In
practice, randomized (normalized) FFT measurements, $\Phi_{\mathcal{B}%
}=W_{\mathcal{B}},(W)_{ij}:=n^{-\frac{1}{2}}e^{-\imath2\pi ij/n}$ for
$i,j=0\dots n-1$, are typically implemented. All performance indicators of the
scheme strongly depend on the size of the control window $\mathcal{B}$ and its
complement $\mathcal{B}^{C}$ where $\mathcal{B}\cup\mathcal{B}^{C}=[n]$. It is
desired to keep the size of the observation window $m\leq|\mathcal{B}|$ as
small as possible to reduce the control overhead. The unused subcarriers in
$\mathcal{B}^{C}$ can then be used to implement further \emph{parallel}
sub-channels for, say, improved user activity detection. In this situation we
shall use $\mathcal{B}_{1}=\mathcal{B}$ and $\mathcal{B}_{2},...,\mathcal{B}%
_{c}\subset\mathcal{B}^{C}$ where $c:=\left\lfloor n/m\right\rfloor >1$ is
some (possibly random) integer. Clearly, signature set and even the channels
(due to different collisions in the sub-channels) require an additional
sub-channel index, i.e. $p^{1},...,p^{c}$ and $h^{1},...,h^{c}$ so that with
$y_{i}^{1},...,y_{i}^{c}$ we have:%
\begin{equation}
y_{i}^{j}=\Phi_{\mathcal{B}}C(p^{j})D_{i}h^{j}+\Phi_{\mathcal{B}}e_{i}%
\end{equation}
Finally, we make the assumptions that there is some kind of load estimation
and power control, which clearly effects the SNR.
We incorporate this into a
normalization on $h^{j}$ though for notational convenience such that:%
\[
\text{\textbf{Power-Control: \ \ }}\mathbb{E}\left\Vert h^{j}\right\Vert
^{2}\leq1
\]
%


Using tailored signature design, of which the details are omitted due to lack of space, we can derive the following proxy measurement model:
\begin{align}
y_{i}^{j}=A_{\mathcal{B}_{j}}D_{i}h^{j}+z_{i},
\label{eq:1stModel}%
\end{align}
where $A_{\mathcal{B}_{j}}$ can be regarded as a randomized subsampled
FFT 
which is normalized by an additional factor of $\sqrt{n/m}$. Under the assumption that the additive noise $e_{i}$ is Gaussian with variance $\sigma^{2}$, we find that
$z\sim\mathcal{CN}(0,\frac{\sigma^{2}}{n}I_{m})$.


\section{Detection algorithms}

\label{sec:algorithms}

The detection algorithm (channel and data) shall be used in parallel on every
sub-channel. For $c>1$ parallel sub-channels, $k_{u}$ splits up to
(random) sub-indices $k_{u}^{1},...,k_{u}^{c}$ where $\sum_{j=1}^{c}k_{u}%
^{j}=k_{u}$. The possibility to reconstruct $\tilde{h}^{j}=\left(  h^{j}%
,D_{1}h^{j},...,D_{t-1}h^{j}\right)  $ from only a small control window
$\mathcal{B}_{j}$ relies on two structural assumptions: The vector $\tilde
{h}^{j}$ containing all CIRs has at most $k_{u}^{j}\cdot k_{s}\cdot t$
non-vanishing entries in total. But the hierarchical structure of the
non-vanishing entries
implies that $\tilde{h}^{j}$ is $(k_{u}^{j},k_{s}%
,t)$-sparse. 

The recovery of $(k_{u}^{j},k_{u}%
,t)$-sparse signals from linear measurements, i.e. h-CS, was studied in
Ref.~\cite{HiHTP, wunder2019low} following the outline of model-based
CS~\cite{BaraniukEtAl:2010}. Therein an efficient algorithm, HiHTP/HiIHT, was
proposed and a recovery guarantee based on the generalised restricted isometry
property (RIP) constants was proven.

Denote the linear (correlation) detector by:
$\Psi_{j}:= (  {A_{\mathcal{B}_{j}}} )  ^{H}$.
Our strategy is to apply hierarchical hard thresholding operator, given by:
\begin{equation}
L_{k_{u},k_{s},t}(x):=\operatorname{supp}\underset{\text{$(k_{u},k_{s}%
,t)$-sparse $y$}}{\operatorname*{arg}\min}\Vert{x-y}\Vert,
\label{eq:threshold_1}%
\end{equation}
which computes the support of the best $(k_{u},k_{s},t)$-sparse approximation,
sub-channel-wise but jointly on $\Psi_{j}y_{0}^{j},..,\Psi_{j}y_{t-1}^{j}$ to
get the support of $h^{j}$, and subsequently to estimate channel and data
on this support denoted $(\mathcal{A}_{j}^{\ast})$
where $(\mathcal{A}_{j}^{\ast})_{k}\neq\{\ \}$ (empty set) if and only if:%
\[
\lVert\left(  h_{k}^{j},D_{1}h_{k}^{j},...,D_{t-1}h_{k}^{j}\right)  \rVert
\geq\sqrt{t\xi}\]
The procedure can be improved by iterate over the residuals of the re-encoded signals
as in HiHTP/HiIHT.

\section{Performance analysis}

\label{sec:performance}

\subsection{Figures of merit}


{\it{Missed detection, false alarm}}. The missed detection probability per user $P_{\text{md}}\left(  \xi\right)  $
is a key metric for the system. Note that by symmetry the probability of a
missed detection is identical for all active users. Since the algorithm works
on each sub-channel in parallel, clearly, $P_{\text{md}}\left(  \xi\right)  $
depends on the load. For $c=1$ the load is simply $k_{u}$. For $c>1$, since an
active user can only appear once in any of the sub-channels so that,
conditioning on a specific sub-channel selection, the resulting $P_{\text{md}%
}(\xi)$ is simply the average over the marginal load distribution in this
sub-channel. Clearly, this is independent of the sub-channel selection and we
could as well just fix a certain sub-channel, say $\mathcal{B}_{1}$, loaded on
average with $\bar{k}_{u}:=1+(k_{u}-1)/c$. Eventually, we define the
probability that some inactive user is falsely detected as active by
$P_{\text{fa}}(\xi)$.



{\it{Collision probability}}. In order to capture the dynamic behavior with parallel $c$
sub-channels, recall that $k_{u}$ out of $u$ devices in total access the
random access channel. The average number of non-collided devices
$E_{\text{nc}}\left(  k_{u}\right)  $ is well-known \cite{dera} and given by:%
\begin{align}
E_{\text{nc}}\left(  k_{u}\right)  
\geq k_{u}-k_{u}\cdot\frac{k_{u}k_{s}}{c\cdot n}%
\end{align}
where the inequality is true for not too large load $k_{u}/u$. Moreover, the
left-hand side is valid for any $k_{u}$, i.e. also for $k_{u}>u$, which would
violate the sparsity constraint though.

Obviously, for $c=1$, there is always a fraction of at least $k_{u}/u$ devices
in collision. For standard CS we have $c\simeq n/k_{u}k_{s}\log({n}/({k}%
_{u}k_{s}))$, it is easily seen that the collision fraction reduces to
approximately $(k_{u}/n)^{2}$ devices. However, we shall see in the following
analysis that this does not capture the real behavior of our system which is
governed by the sparsity capture effect allowing $c=O(n/\log(n))$ and
$r=n/\log^{2}(n)$.


\subsection{User detection analysis}
Assume (for the moment) some minimum channel energy $h_{\min}$ and no noise.
Suppose the energy threshold $\xi$ is chosen as $0<\xi<h_{\min}$ and define
$\epsilon=\min\left\{  \xi,h_{\min}-\xi\right\}  $. We denote the set of all
possible index sets of cardinality $k_{s}$ and indices only in the $k$-th
block by $\Omega_{k}$. The thresholding operator applied to the linear
estimation $\Psi_{j}y_{0}^{j},...,\Psi_{j}y_{t-1}^{j}$ does identify the correct set of users if the following condition holds:
\begin{equation*}
\max_{\substack{k\in\left[  u\right]  \\ \omega\in\Omega_{k}}}\left\vert \sum_{l\in\omega
}\left\vert \left\langle h^{j},v_{l}\right\rangle \right\vert ^{2}-\frac{1}%
{t}\sum_{i\in\lbrack t]}\sum_{l\in\omega}\left\vert \left\langle \Psi_j
y_{i}^{j},v_{l}\right\rangle \right\vert ^{2}\right\vert \leq\epsilon.
\label{eq:e_set}%
\end{equation*}
We start with a new concentration result. 

\begin{theorem}
Let $x$ be fixed with $|\operatorname{supp}\left(  x\right)  |\leq k_{s}k_{u}%
$. Consider terms of the form $\left\Vert AD_{0}x\right\Vert ^{2},\left\Vert
AD_{1}x\right\Vert ^{2},...\left\Vert AD_{t-1}x\right\Vert ^{2}$ where
$D_{0}=I_{n},D_{i},i>0,$ are random unitary diagonal matrices. Then:%
\begin{align*}
&  \mathbb{P}\left(  \left\vert \frac{1}{t}\sum_{i\in\lbrack t]}\left\Vert
AD_{i}x\right\Vert ^{2}-\left\Vert x\right\Vert ^{2}\right\vert >\epsilon
\left\Vert x\right\Vert ^{2}\right)  \\
&  \leq\frac{32\log\left(  2mk_{s}^{2}k_{u}^{2}\right)  +1}{\epsilon^{2}tm}%
\end{align*}

\end{theorem}

The main difference to the standard approach \eqref{eq:Mconc4} is that our approach yields a bound which decays with growing $t$. This is in contrast to typical CS bounds. 
Moreover, a closer inspection yields that it is even true for a statistical
support condition of the form $\mathbb{E}_{k_{u}}|\operatorname{supp}\left(  x\right)  |\leq
k_{s}\bar{k}_{u}$.

We will now exploit this result for the misdetection. Since the analysis is
independent of user index $k$ and sub-channel $j$ we shall set $k=j=1$. For
this let $F^{z}(\xi|x):=\mathbb{P}(\sum_{i}\Vert(h^{1}+z_{i})_{\mathcal{A}%
_{1}}\Vert^{2}\leq\xi|x)$ where $x$ is the (random) sub-channel load
$k_{u}^{1}=x$, i.e. the probability that conditioned on $x$ the energy falls
below some threshold $\xi$. We let the system scale with $n$ and fix a
(non-random sequence) $m_{n}$ from which we get we have $c_{n}=n/m_{n}$ sub-channels.

\begin{theorem}
Let $\epsilon>0$. Let user 1 be in sub-channel 1 with fixed sub-channel load
$x$. There are constants $C_{1},C_{2}$ independent of all parameters:%
\begin{align*}
&  P_{\text{md}}\left(  \xi|x\right)  
  \leq F^{z}(\xi+\epsilon|x)\\
&  +C_{1}u\left(  \frac{es}{k_{s}}\right)  ^{k_{s}}\frac{k_{s}^{2}\log\left(
2mk_{s}^{2}x^{2}\right)  +1}{\epsilon^{2}tm}+C_{2}\left(  \frac
{\operatorname{SNR}}{n}\right)  ^{-k_{s}x}%
\end{align*}

\end{theorem}

Let us interpret the results: First, clearly we see that $P_{\text{md}}\left(
\xi|x\right)  $ under every load is noise-stable. Second, assume that the
sub-channel load $x$ is close to its average $O(k_{u}/c)$ which equals
$O(k_{u}/c)=m$ in the fully loaded system $k_{u}=O(n)$. Since {$h_{\text{min}%
}=O(1/x)$ we require that }$\epsilon={O(1/x)}$. Hence, if we set
$m=x=O(\log(n))$, $u=O(n/x^{2})$ and $t=n$ we see that $P_{\text{md}}\left(
\xi|x\right)  \rightarrow0$ (fixed $k_{s}$). Hence, for the result to hold we
need to show that each sub-channel load is indeed not higher than
$O(\log\left(  n\right)  )$ with high probability so that in each sub-channel
the collision turns to zero.

\subsection{User collision analysis}

The sparsity capture effect works as follows: We can think of the baseline
system as distributing $k_{u}$ users for the $n$ resources (in the frequency
domain) which leads to collisions. Now suppose we bundle $y\left(  x\right)  $
resources together, call it the sub-channel, and possess a mechanism that can
handle $x$ users with high probability (i.e., by our hierarchical CS sensing
algorithm). The resulting number of sub-channels as $c\left(
x\right)  =n/y\left(  x\right)  $. Clearly, such sub-channel load is
binomial-distributed. 
By applying the Poisson approximation (which is exact for large $n$) and the union
bound, 
the following result shows indeed that the number of subchannels $c\left(
x\right)$ grows fast enough to ensure that the effective load in the sub-channels remains sparse with overwhelming probability. 
\begin{theorem}
(sparsity capture effect) If $y(x)\leq x$ then:%
\begin{equation*}
\mathbb{P}\left(  \{x\geq k_u^1\}\right)
 \leq\frac{n\left(  k_{u}-x\right)  }{x\sqrt{2\pi x}}e^{-\left(
1-\frac{k_{u}}{n}\right)  ^{2}x}%
\end{equation*}

\end{theorem}

Let us again interpret the result: If the number of measurements equals
$m=O(\log(n))$, then the probability that $x\geq O(((1-\frac{k_{u}}{n}%
)^{-2}\log n))$ or more users are in some specific sub-channel turns to zero
for any load $k_{u}/n<1$ as $n$ grows. Hence sub-channel load $x$ concentrates around
$O(\log(n))\ $ 
and probability of collision in each sub-channel turns to zero with $n$ as well, obviously, thus completing our analysis. We will
validate now our results in the next section.

\section{Evaluations and Simulations}

\label{sec:evaluation}
Parallel sub-channels are
created by randomly partitioning the $n$-dimensional image space into blocks
of length $m$, leading to $c=n/m$ sub-channels. For each sub-channel
$j=1,\ldots,c$, a vector $x^{j}\in\mathbb{C}^{n}$ is divided into $u$ blocks,
each of length $s$, such that $n=us$. If a user chooses resource block
$k\in\lbrack u]$, block $x_{k}$ is filled with the $k$th user's $k_{s}$-sparse
signature. We allow for $k_{u}^j=\bar{k}_u$ users per sub-channel. Each user is also
encoding data into diagonal matrices $D_{i}$, $i=2,\ldots,t$ containing
entries of modulus 1. Hence, at the BS data blocks $y_{i}^{j}=A_{j}D_{i}%
x^{j}\in\mathbb{C}^{m}$ for $j=1,\ldots,c,i=0,\ldots,t$ are received, forming
the observation $y\in\mathbb{C}^{c\times m\times t}$. Here, $A_{j}$ is a
matrix consisting of $m$ rows of a $n\times n$ DFT matrix, corresponding to
the frequencies allocated to sub-channel $j$.

\begin{figure}[ptb]
\centering
\includegraphics[width=.45\textwidth]{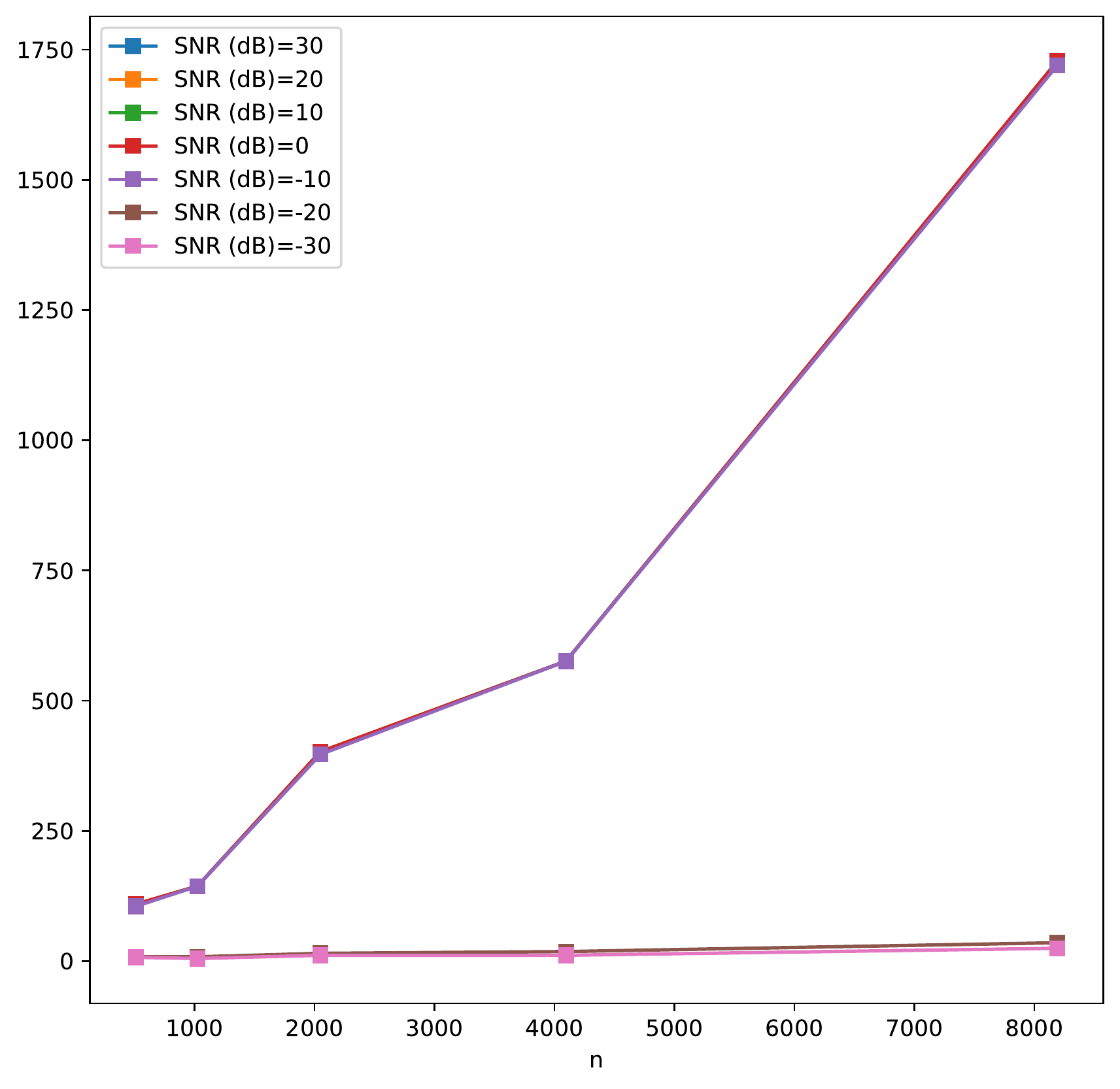} \caption{Average
number of supported users (100 Monte-Carlo trials)}%
\label{fig:supported_users}%
\end{figure}

User detection is performed by one step of Hi-IHT \cite{wunder2019low}.
The number $\bar{k}_u$ of users per sub-channel is chosen such that the
probability of 2 or more users trying to access the same resource block is
below a preset probability $0<p_{u}<1$, i.e. the largest $k\in\mathbb{N}$ such
that
\begin{equation}
\prod\limits_{i=1}^{k}\left(  1-\frac{i}{n}\right)  \geq1-p_{u}%
.\label{eq:select_ku}%
\end{equation}
The left hand side of inequality \eqref{eq:select_ku} is the probability that
each of $k$ indices out of $[u]$ selected uniformly at random are unique. 
Hence, on average the total number of supported users is given
by $(1-p_{u})\cdot \bar{k}_u\cdot c\cdot (1-P_{md})$.
In our simulations we
considered blocks of length $s=8$ with an in-block sparsity $k_{s}=4$. Then,
$\bar{k}_{u}$ was selected such that $p_{u}\geq0.1$. We
set $t=100$ and $m=2^{\lfloor\log_{2}(\bar{k}_{u}\cdot k_s)\rfloor}$, which resulted
in detection rates close to 1 in noise-free simulations. The number of
supported users in this setting for $n=2^{10},\ldots,2^{13}$ can be observed
in fig \ref{fig:supported_users}, which shows that the system also performs
well under noise. With a SNR $\geq-10$dB the system performance is virtually
indistinguishable from the noise free case. 
To give concrete numbers: Assuming 60kHz subcarrier spacing and FFT size 4096 the system reliably detects 450000 devices
per second with good load, e.g. estimated 100kByte for rate $1/4$ code. 

\section{Conclusion}
Exploiting a new measure concentration inequality, we designed a massive random access scheme based on hierarchical compressed sensing, conducted theoretical performance analysis and demonstrated its feasibility by numerical simulations. The proposed scheme promises huge gains in terms of throughput and number of supported users. 

\bibliographystyle{unsrt}
\bibliography{sspa}

\begin{thebibliography}{10}

\bibitem{Wunder2015_ACCESS}
G.~Wunder, H.~Boche, T.~Strohmer, and P.~Jung.
\newblock {Sparse Signal Processing Concepts for Efficient 5G System Design}.
\newblock {\em IEEE ACCESS}, December 2015.

\bibitem{Bockelm2018_ACCESS}
C.~Bockermann, N.~Patras, G.~Wunder, and et~al.
\newblock {Towards Massive Connectivity Support for Scalable mMTC
  Communications in 5G networks}.
\newblock {\em IEEE ACCESS}, May 2018.

\bibitem{Rahut2013}
Simon Foucart and Holger Rauhut.
\newblock {\em A mathematical introduction to Compressed Sensing}.
\newblock Birkh\"auser, 2013.

\bibitem{Choi2018_TVC}
J.~{Choi}.
\newblock Stability and throughput of random access with cs-based mud for mtc.
\newblock {\em IEEE Transactions on Vehicular Technology}, 67(3):2607--2616,
  2018.

\bibitem{Choi2020_IoT}
J.~{Choi}.
\newblock On throughput of compressive random access for one short message
  delivery in iot.
\newblock {\em IEEE Internet of Things Journal}, 7(4):3499--3508, 2020.

\bibitem{Wunder2015_GC}
Gerhard Wunder, Peter Jung, and Mohammed Ramadan.
\newblock {Compressive Random Access Using A Common Overloaded Control
  Channel}.
\newblock In {\em IEEE Global Communications Conference (Globecom'14) --
  Workshop on 5G \& Beyond}, San Diego, USA, December 2015.

\bibitem{dera}
M.~I. {Hossain}, A.~{Azari}, and J.~{Zander}.
\newblock Dera: Augmented random access for cellular networks with dense
  h2h-mtc mixed traffic.
\newblock In {\em 2016 IEEE Globecom Workshops (GC Wkshps)}, pages 1--7, Dec
  2016.

\bibitem{HiHTP}
I.~{Roth}, M.~{Kliesch}, A.~{Flinth}, G.~{Wunder}, and J.~{Eisert}.
\newblock Reliable recovery of hierarchically sparse signals for {G}aussian and
  {K}ronecker product measurements.
\newblock {\em IEEE Transactions on Signal Processing}, 68:4002--4016, 2020.

\bibitem{wunder2019low}
Gerhard Wunder, Stelios Stefanatos, Axel Flinth, Ingo Roth, and Giuseppe Caire.
\newblock Low-overhead hierarchically-sparse channel estimation for multiuser
  wideband massive mimo.
\newblock {\em IEEE Transactions on Wireless Communications}, 18(4):2186--2199,
  2019.

\bibitem{BaraniukEtAl:2010}
R.~G. Baraniuk, V.~Cevher, M.~F. Duarte, and C.~Hegde.
\newblock Model-based compressive sensing.
\newblock {\em IEEE Trans. Inf. Th.}, 56(4):1982--2001, April 2010.

\end{thebibliography}

\end{document}